\documentclass[%
 reprint,
 amsmath,amssymb,
 aps,
]{revtex4-1}
\usepackage{caption}
\usepackage[english]{babel}
\usepackage[utf8x]{inputenc}
\usepackage{color}
\usepackage{graphicx}
\usepackage{dcolumn}
\usepackage{bm}
\usepackage{subcaption}
\usepackage{comment}
\usepackage{mathtools}
\usepackage{cleveref}

\newcommand{\IT}[1]{{\color{blue}#1}}

\usepackage{physics}

\begin{document}


\title{Optical Bloch modeling of few-cycle laser-induced electron dynamics in dielectrics}
\author{E. Smetanina$^{1,2}$}
\author{P. Gonz\'{a}lez de Alaiza Mart\'{\i}nez$^{1}$}
\author{I. Thiele$^{3}$}
\author{B. Chimier$^{1}$}%
\author{A. Bourgeade$^{1}$}%
\author{G. Duchateau$^{1}$}
 
\affiliation{%
$^{1}$ University of Bordeaux-CNRS-CEA, Centre Lasers Intenses et Applications, UMR5107, 33405 Talence, France
}
\affiliation{%
$^{2}$ Department of Physics, University of Gothenburg, SE-412 96 G{\"o}teborg, Sweden
}
\affiliation{%
$^{3}$ Department of Physics, Chalmers University of Technology, SE-412 96 G{\"o}teborg, Sweden %
}

\date{\today}

\begin{abstract}
We develop a new model of laser-matter interaction based on Optical Bloch Equations, which includes photo-ionization, impact ionization, and various relaxation processes typical of dielectric materials. This approach is able to describe the temporal evolution of the electron dynamics in the conduction band driven by few-cycle laser pulses of any wavelength. Moreover, the nonlinear polarization response of both centrosymmetric and non-centrosymmetric materials can be described while ensuring the proper selection rules for the harmonics emission. 
\begin{description}
\item[PACS numbers]
 52.50.Jm, 52.25.Tx, 79.20.Ds,42.65.Ky
\end{description}
\end{abstract}

\pacs{Valid PACS appear here}
\maketitle


\section{\label{sec:Intro}Introduction}
Modern laser technologies provide high-intensity few-cycle laser pulses which open new doors for studies of laser-matter interaction processes. In case of semiconductors or dielectric material targets, such laser pulses can be used to drive the electronic populations in excited states, then allowing various applications. 
For instance, an initially insulating dielectric material can be reversibly driven to a conducting state within a femtosecond temporal resolution \cite{Schultze_Nature_2013_Controlling_dielectrics_by_light}. 
The interaction of such pulses with dielectric solids can also lead to high order harmonic generation and THz radiation \cite{You2017,HohenleutnerNature2015,SchubertNatPhot2014}, providing applicative capabilities of a strong interest~\cite{Kampfrath13}. 
The benefit of ultrashort femtosecond pulses is also important in fields more close to industrial applications as material ablation, surface texturing and others \cite{Itina_Uteza2009,Chimier_PhysRevB.84.094104,ITINA_SciRep_2017,Courvoisier_Meyer:17, Courvoisier_doi:10.1063/1.5053085,Balling_W_degaard_2014_ablation, Balling_2013_thin_film_ablation}. In that case, the relaxation of the energy of the excited electrons towards the lattice through collisions leads to an energy deposition into the material \cite{TwoTemperatures, Bulgakova_JAppPhys2015,Bulgakova_SciRep2016}. The possibility to improve the efficiency of previous applications has also been demonstrated by designing the pulse characteristics including a broad spectrum or a spatio-temporal chirp \cite{ Gulley_chirp_PhysRevB.90.155119, Kazansky_SciRep_2017, Arabanian:14, Vitek:10, Geltikov_SciRep_2018, Ultrafast_Ablarion,Ultrafast_Ablation_Pasquier:17}. In order to interpret the experimental results, understand the physical mechanisms at play, and guide the experimental developments for further studies, it clearly appears that time-dependent models for the laser induced electron dynamics by few-cycle pulses in dielectric solids are required. In view of the above-mentioned studies, such a model should account for the photo-ionization and the electron dynamics in the conduction band where electrons may further absorb photons and undergo collisions. Ultimately, these models should be suitable for their implementation into a Maxwell solver \cite{BidegarayFesquet2006,Bidegaray2001} or Unidirectional pulse propagation equation solver \cite{Kolesik2002,Kolesik2004} to describe the coupled laser propagation and electron dynamics that generally take place for such physical systems \cite{BOURGEADE2006823, Maxwell-Bloch_SAUT2005927, Maxwell_Bloch_M2AN_2004__38_2_321_0, Duchateau_PhysRevA.89.053837, Bourgeade2012,Gulley_PhysRevB.90.155119, Kolesik:18}.

Various classes of such models have been developed. Going from the crudest to the first principle approaches, the main models are as follows. A single rate equation for the evolution of the electron density in the conduction band has been shown to provide global observed trends as the electron avalanche due to impact ionization. This approach has been improved by multiple rate equations (MRE) accounting partly for the band structure \cite{Rethfeld_PhysRevLett.92.187401, Balling_MREs_PhysRevB.79.155424, Gallais2015}. However optical cycle-averaged transition rates still stand. This assumption has been overcome by solving optical Bloch equations (OBEs) where the time-dependent laser electric field is the input parameter~\cite{Gulley15}. Nevertheless, the impact ionization process was not tackled, neither an in-depth treatment of collisions. On the other hand, kinetic-type descriptions have been developed, accounting for all main collisional processes \cite{Kaiser_PhysRevB.61.11437, Barilleau_2016, Gruzdev:17}. The laser pulse intensity-induced evolution of the band curvature is considered to be responsible for the material breakdown in \cite{Gruzdev2008, Gulley_PhysRevB.90.155119, Gruzdev_2018_PhysRevB.98.115202}. But they generally do not account for the band structure beyond a single parabolic band and are currently computationally too expensive for their coupling to a Maxwell solver. First principle approaches as time-dependent density functional theory fully describe the band structure, as well as the time-dependent interaction \cite{Yakovlev_prop_evol, Yabana_PhysRevB.92.205413}. However collisions are not well described \cite{Otobe_doi:10.1063/1.4864662, Otobe_doi:10.1063/1.4867438} whereas their influence is crucial since they lead to decoherence effects and to the laser energy deposition into the material. To our knowledge, there is no model including the following required features: time-dependent laser electric field as input of the model, description of the band structure, description of collisions, photo-ionization, electron heating in the conduction band, description of impact ionization, and electron recombination.

In the present work, such a model is proposed based on the optical Bloch equations which consists in solving the Liouville equation for the density matrix (Sec.~\ref{sec:MODEL}). A band structure is introduced through a set of energy levels. By coupling appropriately those levels, photo-ionization, impact ionization, transitions in the conduction band, and electron recombination are taken into consideration. 
The solution of these equations for the density matrix allows to determine both, the electron population for each energy level and the corresponding ionization rates as well as the polarization including its linear and nonlinear part, for both centrosymmetric and non-centrosymmetric dielectric materials (Sec.~\ref{sec:Results}). The electron population determines the energy density of the whole electron gas and is useful for the calculation of the absorbed energy density in the material. The polarization gives access to the description of secondary radiation, including harmonics and terahertz (THz) radiation. 
\par
This work aims at providing a theoretical baseline to address a consistent time-dependent modeling of the electron dynamics induced by ultra-short few-cycle laser pulses, where all main collisional processes taking place in dielectric materials are considered. The reliability of this approach is provided by the above-mentioned cases of interest where standard trends are retrieved. Since the polarization can be easily extracted from the density matrix, the present Bloch approach is very suitable for coupling to Maxwell solvers \cite{BidegarayFesquet2006,Bidegaray2001}, thus providing a route to model accurately the complex electron-laser propagation dynamics which is required to account for experimental conditions.

\section{\label{sec:MODEL}Model}


The OBEs are constructed from the Liouville-von-Neumann equation which describes the time evolution of a quantum system using the density matrix formalism. To include impact ionization and various relaxation processes taking place in dielectric materials, the equation for density matrix $\hat{\rho}$ evolution reads \cite{Bidegaray2001,Bourgeade2012}:
\begin{equation}\label{eq:MODEL:MAIN}
\partial_t{\hat{\rho}} = \hat{\cal L}\left(\hat{\rho} \right) + \hat{\cal G}_{\rm r}\left(\hat{\rho} \right)  + \hat{\cal G}_{\rm imp}\left(\hat{\rho} \right), 
\end{equation}
where $\hat{\cal L}\left(\hat{\rho} \right)$ is the so-called Liouville-von-Neumann super-operator, and the super-operators $\hat{\cal G}_{\rm r}\left(\hat{\rho} \right)$ and $\hat{\cal G}_{\rm imp}\left(\hat{\rho} \right)$ introduce phenomenological relaxation (e.g., recombination and coherence loss) and impact ionization terms, respectively. The numerical scheme developed to solve the present optical Bloch equations is provided in the Appendix~\ref{sec:appendix:numerics} .

\subsection{\label{subsec:MODEL:FIELD_IONIZATION}Liouville-von-Neumann super-operator}

The Liouville-von-Neumann super-operator $\hat{\cal L}\left(  \hat{\rho} \right)$ is given by \cite{Bidegaray2001}:
\begin{equation}\label{eq:model:LiouvillevonNeumannEquation:L}
\hat{\cal L}\left(  \hat{\rho} \right) := -\frac{\rm i}{\hbar} \,\comm{\hat{H}}{\hat{\rho}} = -\frac{\rm i}{\hbar} \left(  \hat{H}\hat{\rho} - \hat{\rho} \hat{H}  \right),
\end{equation}
where $\hat{H}(t)$ is the electron Hamiltonian and reads:
\begin{equation}\label{eq:model:H}
\hat{H}(t) = \hat{H_0} + \hat{V}(t).
\end{equation}


The unperturbed Hamiltonian $\hat{H_0}$ is modeled as a diagonal matrix including considered energy levels. We consider a finite number $N$ of allowed energy levels in the CB. The corresponding indexes for the CB level indication are $0 < j \leq N$. 
Depending on the underlying material properties~(see Sec.~\ref{subsec:MODEL:SYMMETRY} for more details), the valence band~(VB) contains one single level with the index $i = 0$ or two energy levels with almost the same energy which have the indexes $ j = -1$ and $ j = 0$ . For example, in case of two VB states the Hamiltonian reads:
\begin{equation}\label{eq:model:OBEs:H0}
\hat{H_0} = \mqty[\dmat[0]{E_{-1},E_0,E_1,\ddots,E_N}].
\end{equation}

The levels in the CB have been chosen to have the following energies \cite{Rethfeld_PhysRevLett.92.187401,Rethfeld_PhysRevB.73.035101}:
\begin{equation}\label{eq:model:Ej}
E_j = E_g + (j-1)\hbar\omega_0,
\end{equation}
where $\hbar\omega_0$ is the photon energy of the incident light and $E_g$ is the gap energy of the considered material. The highest considered CB level $j=N$ must have an energy $E_N-E_1$ that is sufficiently large to fulfill energy and momentum conservation during the impact ionization process \cite{Rethfeld_review}:
\begin{equation}\label{eq:model:EN}
E_N - E_1 \geq 1.5 \, E_g,
\end{equation}
and the number of CB levels that will be considered is:
\begin{equation}\label{eq:model:N}
N   = 1 + \left\lfloor \frac{ 1.5 \, E_g}{\hbar\omega_0} \right\rfloor,
\end{equation}
where $\lfloor x \rfloor$ is the floor function (maximum integer that is less or equal than $x$). This estimation assumes that the electron masses in VB and CB are equal to the free-electron mass and neglect the mean oscillation energy of the applied electric field~\cite{Rethfeld_PhysRevLett.92.187401}. 

The interaction Hamiltonian $\hat{V}$ in Eq.~\eqref{eq:MODEL:MAIN} is calculated with the dipole approximation in the length gauge and in one-dimensional geometry (i.e., $\vec{r}=y\,\vec{\rm e}_y$) as follows \cite{CohenTannoudji1998_I,CohenTannoudji2000_II}:
\begin{equation}\label{eq:model:OBEs:V}
\hat{V}(t) = - e E(t) \hat{\mu},
\end{equation}
where $e$ is the elementary charge, $E(t)$ is the instantaneous value of the laser electric field and $\hat{\mu}$ is the dipole transition matrix, which is Hermitian.

\subsection{\label{subsec:MODEL:Relaxation} Relaxation super-operator}

In this paper we consider two relaxation processes, namely, recombination ($ \hat{\cal G}_{\rm rec} $) and relaxation of coherence ($ \hat{\cal G}_{\rm coh} $):
\begin{equation}\label{eq:model:introduce G:G = rec + coh}
 \hat{\cal G}_{\rm r}\left(\hat{\rho} \right) = \hat{\cal G}_{\rm rec}\left(\hat{\rho} \right) + \hat{\cal G}_{\rm coh}\left(\hat{\rho} \right).
\end{equation}

The electron recombination process is introduced as a decay of CB electrons to VB on the timescale of $\tau_{\rm rec}$. The equations describing the recombination process in case of one VB level are as follows:
\begin{equation}\label{eq:model:G rec i}
\hat{\cal G}_{\rm rec}\left(\hat{\rho} \right) \equiv \left \lbrace  \begin{array}{ll}
     \displaystyle \partial_t \rho_{j,j} = - \rho_{j,j}/ \tau_{\rm rec}, & \text{for $ 1 \leq j \leq N$,} \\
      \\
    \displaystyle  \displaystyle \partial_t \rho_{0,0} =  \sum_{j=1}^N  \rho_{j,j}/ \tau_{\rm rec} ,
\end{array}   \right .
\end{equation}
where $\tau_{\rm rec}$ is the characteristic recombination time. We consider a recombination time of $\tau_{\rm rec} \sim 150$~fs for fused silica glass \cite{Audebert1994, Tzortzakis2001, Rolle2014}. If there are two VB levels, electrons recombine indifferently to any of these two VB levels, provided that the energy separation between those VB levels is negligible.

The coherence loss introduces the dissipation of electron energy to the lattice due to electron-phonon collisions and is modeled by an exponential decay of the off-diagonal density matrix elements:
\begin{equation}\label{eq:model:G coh}
\hat{\cal G}_{\rm coh}\left(\hat{\rho} \right) \equiv \left \lbrace  \begin{array}{ll}
     \displaystyle \partial_t \rho_{j,k} = - \rho_{j,k}/ \tau_{\rm coh}, & \text{for $j\neq k$,} \\
\end{array}   \right .
\end{equation}
where $\tau_{\rm coh}$ is the coherence-loss characteristic timescale. We take $\tau_{\rm coh} \sim 1\mbox{--}10$~fs for coherence relaxation driven by electron-phonon collisions 
\cite{Arnold1992}.


\subsection{\label{subsec:MODEL:ImpIonSuperOp} Impact-ionization super-operator}

The impact ionization is introduced by the impact-ionization super-operator $\hat{\cal G}_{\rm imp}$ and describes how an electron in the highest CB level, by loosing energy, promotes the ionization of an electron of a VB level and both electrons go into the lowest CB level. The corresponding equations describing impact ionization in case of one VB level read as follows:
\begin{equation}\label{eq:model:G imp (k)}
\hat{\cal G}_{\rm imp}\left(\hat{\rho} \right) \equiv \left \lbrace  \begin{array}{l}
     \displaystyle \partial_t \rho_{N,N} = - \rho_{N,N}/ \tau_{\rm imp},   \\
      \\
     \displaystyle \partial_t \rho_{1,1} = 2\rho_{N,N}/ \tau_{\rm imp},   \\
      \\
    \displaystyle  \displaystyle \partial_t \rho_{0,0} =  - \rho_{N,N}/ \tau_{\rm imp} ,
\end{array}   \right .
\end{equation}
where $\tau_{\rm imp}$ is the characteristic impact-ionization timescale. The characteristic timescale of impact ionization for fused silica glass is taken $\tau_{\rm imp} \sim 1$~fs \cite{Kruchinin}. In case of a two VB-levels, the impact ionization process is modeled separately for the levels 0, 1 and N and for the levels -1, 1 and N with the same time scale. Implicitly, we require that the energy of the highest CB state is sufficiently large in order to fulfill energy and momentum conservation during the impact ionization process \cite{Rethfeld_review}: $E_N - E_1 \geq 1.5 \, E_g$.


\section{\label{sec:Results} Results and Discussion}


Our OBEs model will be used to study the tme evolution of the electron dynamics as well as the all-order polarization response of dielectric materials to few-cycle laser pulses. Two different material types, with similar band gap values, will be considered: centrosymmetric (like fused silica) and non-centrosymmetric (like crystalline dielectrics \cite{Impact}) materials. They will have similar ionization rates induced by femtosecond high-intensity laser pulses, but completely different polarization responses. For the former materials the nonlinear polarization response contains only odd harmonics, while for the latter all harmonics may be present.

\subsection{\label{subsec:MODEL:SYMMETRY}Material modeling}

Materials with band gap $E_g=9$~eV and a density of neutral atoms $N_0 = 2.2\times10^{22}$~cm$^{-3}$, close to the fused silica values, are considered~\cite{Wu2005}. 
For the centrosymmetric material, we consider two VB levels, namely $j = -1$ and $j=0$, separated by an energy $\Delta E_{-1,0} \ll E_g$. The wave-functions associated to energy levels will have a well-defined parity alternating from level to level and hence $\mu_{j,j} = 0$. Both the elements $ \mu_{-1,0} = \mu_{\rm VB} $ and $\mu_{0,1}$ are free parameters. The latter defines the transitions between VB and CB levels, allowed in the following way: 
\begin{equation}\label{eq:model:mu_above_threshold_centrosymmetric_odd}
\mu_{l,j>1} = \mu_{j>1,l} = \mu_{0,1}\frac{E_1-E_0}{E_j-E_0},
\end{equation}
where $l=0$ if $j$ is odd and $l=-1$ if $j$ is even.
%
Moreover, the transitions between CB levels, representing the laser-induced electron heating process, are allowed between states with different parities and modeled by transition-dipole matrix elements inversely proportional to the energy difference between the levels:
\begin{equation}\label{eq:model:mu_CB_centrosymmetric}
\mu_{j,k}= \mu_{k,j} =\mu_{\rm CB} \frac{1}{|E_j-E_k|},
\end{equation}
%
where $j>0$, $k>0$, $j = k \pm 1,3,5,7...$, and $\mu_{\rm CB}$ is a free parameter defining transition matrix elements connecting the CB levels.

For the non-centrosymmetric-material model, we consider only one VB level with a wave-function that does not have a well-defined parity and the corresponding matrix element is non-zero: we take $\mu_{0,0} = \mu_{\rm VB}$. The transitions from VB level are allowed to any CB level in this case and corresponding matrix elements are given by Eq.~\eqref{eq:model:mu_above_threshold_centrosymmetric_odd} for $l=0$ and every value of $j$. 
In the CB, similarly to the previous case, the wave-functions have well-defined parity that is alternating from level to level. The matrix elements for transitions between CB levels are given by Eq.~\eqref{eq:model:mu_CB_centrosymmetric} for $j>0$, $k>0$, $j = k \pm 1,3,5,7...$. 
This configuration provides both odd- and even-harmonic polarization response and allows us to describe non-centrosymmetric wide-gap dielectrics ionization dynamics.

The VB dipole-transition-matrix element $\mu_{\rm VB}$ is set to 2~\AA. The matrix element $\mu_{0,1}=\mu_{1,0}$ for transitions between VB and CB is set to 0.5~\AA. The CB dipole-transition-matrix element $\mu_{\rm CB}$ is set to 0.45~eV$\cdot$\AA. For the centrosymmetric material model the VB level separation is $\Delta E_{-1,0} = 0.01$~eV. Small variations of these parameters do not lead to any strong fluctuation in the obtained results. These dipole-transition matrix elements have been chosen to closely reach the ionization degree provided by the Keldysh theory at a photon energy equal to 1.5~eV \cite{Keldysh1964}.



\subsection{\label{sec:Results:Parameters} Laser modeling}

The electric laser field is defined in the time interval $0 \leq t \leq \tau_0$ by
\begin{equation}\label{eq:model:laser}
E(t) = {\cal E}_0 \sin^2\left(\pi t/\tau_0\right) \sin\left(\omega_0t\right),
\end{equation}
where $\omega_0$, $\tau_0$ and ${\cal E}_0$ are the central angular frequency, the duration and the amplitude of the laser pulse, respectively. In the following, the duration is expressed in terms of the number of cycles:
\begin{equation}\label{eq:model:laser:numbercycles}
N\text{cycles}=\frac{\tau_0 \omega_0}{2\pi} ,
\end{equation}
and we focus on examples with 5 and 10 cycles. The laser intensity is defined as $I_0 = n_0 \varepsilon_0 c \, {\cal E}_0^2 /2$, where $c$ is the speed of light, $\varepsilon_0$ is the vacuum permittivity and $n_0$ is the refractive index that is set to $1.5$.

We consider laser pulses with $\tau_0$ equal to 5 
cycles, photon energy $\hbar\omega_0$ varying from 0.5~eV to 3~eV (i.e., from the mid-infrared to the ultraviolet spectral regions), and peak laser-pulse intensities $I_0$ going from $10^{9}$~W/cm$^{2}$ up to $5\times10^{14}$~W/cm$^{2}$.
\subsection{\label{sec:III:ionization} Ionization rate}

\begin{figure}[ht]
\centering
\includegraphics[width=0.85\linewidth]{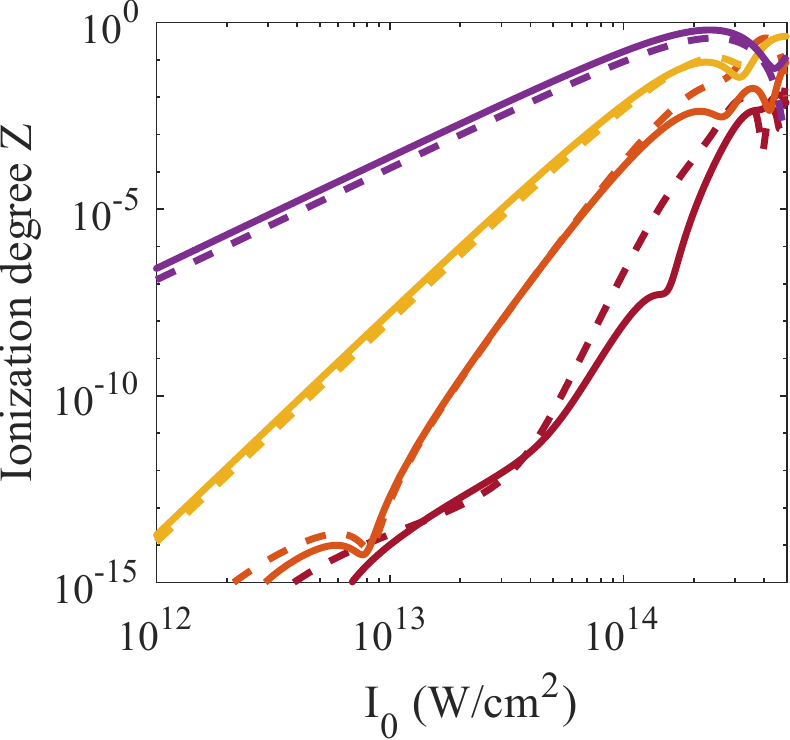}
\caption{\label{fig:Ionization_fig} Ionization degree as a function of the 5-cycle laser intensity for non-centrosymmetric material (solid curves) and centrosymmetric material (dashed curves), for several photon energies: $\hbar\omega_0=3.0$~eV (violet), 1.5~eV (yellow), 1.0~eV (orange), 0.6~eV (red).}
\end{figure}

The CB electron density is estimated as the probability of finding an electron in the CB multiplied by the density of neutrals $N_0$:
\begin{equation}\label{eq:model:Ne}
N_{\rm e}(t) = N_0 \, \sum_{j>0}^N \rho_{j,j}(t) = N_0 \left(1-\sum_{j \leq 0}^0 \rho_{j,j}(t)\right) .
\end{equation}

The unperturbed state corresponds to the electron being in the VB. If the VB is associated with one energy level $j=0$, the initial condition is $\rho_{0,0}(t=0)=1$. If the VB is associated with two closely laying energy levels $j=-1$ and $j=0$, the levels are suggested to be equally populated and the corresponding initial conditions are $\rho_{-1,-1}(t=0)=\rho_{0,0}(t=0)=1/2$.  

The ionization rate in the OBEs model is given by the variation of the electron population in the CB levels over time:
\begin{equation}\label{eq:model:OBEs:W}
W(t) = N_0 \sum_{j > 0} \partial_t \rho_{j,j}(t) = \partial_t N_{\rm e}(t),
\end{equation}
and the ionization degree is defined as the electron density at the final instant $t=\tau_0$ normalized by the density of neutral atoms $N_0$:
\begin{equation}\label{eq:model:iondegree}
Z = \frac{N_{\rm e}(\tau_0)}{N_0} = \sum_{j > 0}^N \rho_{j,j}(\tau_0).
\end{equation}

In order to study numerically the ionization given by our OBEs model, we consider here a simplified version of the systems associated to centrosymmetric and non-centrosymmetric materials having only one CB level (i.e., $j=1$). 
The relaxation and impact-ionization super-operators are excluded.
\begin{figure}[ht!]
\centering
\includegraphics[width=0.85\linewidth]{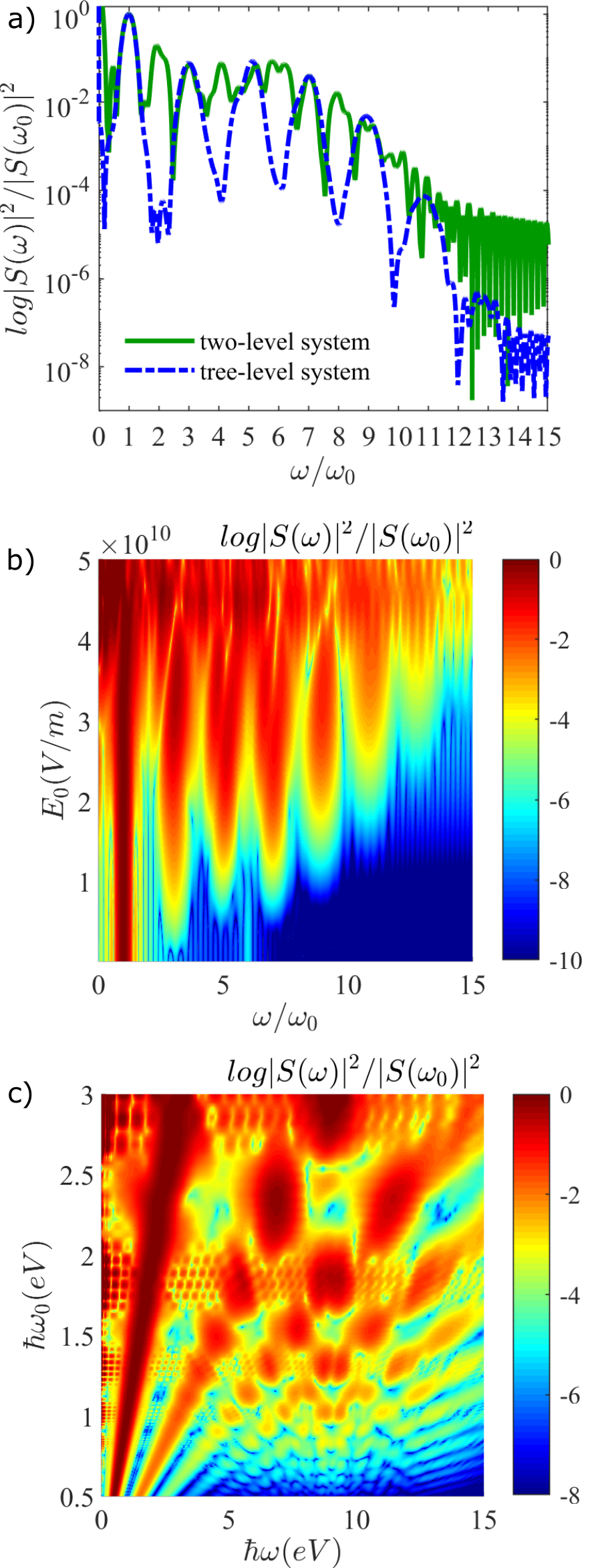}
\caption{\label{fig:Polarization_Fig} Polarization spectrum calculated for 5-cycle laser pulses. (a) Polarization spectrum is calculated for two-level (green curves) and three-level (blue curves) systems at $\hbar\omega_0 = 1.5$~eV and $I_0 = 1.2\times10^{14}$~W/cm$^{2}$. (b) Polarization spectrum for three-level system at $\hbar\omega_0 = 1.5$eV and $E_0$ from 1~GV/m to 50~GV/m. (c) Polarization spectrum is calculated for three-level system at $I_0 = 1.2\times10^{14}$~W/cm$^{2}$ and $\hbar\omega_0$ from 0.5~eV to 3~eV.}
\end{figure}


Figure~\ref{fig:Ionization_fig} presents the ionization degree as a function of the laser intensity for centrosymmetric and non-centrosymmetric materials and  
for several photon energies. The  multi-photon behavior (i.e., $Z\propto I_0^K$ where $K$ is the number of absorbed photons) is found until the system saturates and further oscillations of the ionization degree appear for $I_0 > 2\times10^{14}$~W/cm$^{2}$.

The multi-photon order $K=3$ for $\hbar \omega_0=3$~eV and $K=6$ for $\hbar \omega_0=1.5$~eV, which corresponds to $K=E_g/(\hbar \omega_0)$. In contrast, for lower photons energies, the broad spectrum due to the short laser duration makes the multi-photon order be intensity-dependent because the contribution of higher spectral frequencies is less negligible.

Note that in this work we have neglected in Eq.~\eqref{eq:MODEL:MAIN} the term on $\grad {\bf k}$, where ${\bf k}$ is the wave-vector along the BZ \cite{Gulley15,Sato_intraband_PhysRevB.98.035202,Kruchinin}. Due to this simplification, the model reproduce reproduces only the vertical multi-photon electron transition (resonant case).


\subsection{\label{sec:III:polarization} Polarization response of a single-level CB system}
We compute the all-order polarization response including both linear and nonlinear components as the expectation value of the electric dipole moment $e \hat{\mu}$ in the electron subsystem multiplied by the density of neutrals $N_0$:
\begin{equation}\label{eq:model:OBEs:P}
P(t) = N_0  \Tr\left\lbrace e\hat{\mu}\,\left(\hat{\rho} -\hat{\rho}(0) \right) \right\rbrace,
\end{equation}
where we impose the initial condition $P(t < 0)=0$ assuring the causality of the model \cite{Peterson1973,Hu1989}.

In Fig. \ref{fig:Polarization_Fig}, we utilize the normalized polarization power spectrum defined as: 
\begin{equation}\label{eq:model:polarization:S}
S(\omega) = \frac{\left| {\cal F}\lbrace P(t) \rbrace(\omega) \right|^2} {\left| {\cal F}\lbrace P(t) \rbrace(\omega_0) \right|^2 },
\end{equation}
where ${\cal F}\lbrace P(t) \rbrace(\omega)$ accounts for Fourier transformation.
The polarization response obtained for the two-level system having wave-functions without a well-defined parity is fundamentally different from the polarization response of a three-level system having wave-functions with a well-defined parity, as illustrated by Fig. \ref{fig:Polarization_Fig}(a). 
In the three-level system modeling a centrosymmetric medium  the spectrum of the polarization response has only odd harmonics 1, 3, 5... 
Instead, in the two-level system for non-centrosymmetric media, the spectrum has both even and odd harmonics.

Figure~\ref{fig:Polarization_Fig}(b) reveals that the harmonics cut-off has a linear dependence on the incident electric field amplitude $E_0$, which agrees with other works where the OBEs were used for HHG simulations \cite{HighharmonicgenerationfromBlochelectronsinsolids} and experiments~\cite{High-Harmonics-solids}. Particularly, as in \cite{High-Harmonics-solids}, the harmonics spectrum has well-pronounced maximum at the band gap energy $E_g$ and a linear cut-off dependence for harmonics with photon energies $\hbar\omega$ above $E_g$. In general, the structure of the level system affects the HHG cut-off, i.e., in case of highly multilevel systems several cut-offs in the HH spectrum can be observed \cite{High-Harmonics-solids2,High-Harmonics-semiconductors2,Multilevel_HHG_coupled_pairs_CLs}. 

Figure~\ref{fig:Polarization_Fig}(c) shows the harmonics power-spectra generated at various incident photon energies at fix laser pulse intensity $1.2\times10^{14}$~W/cm$^{2}$. 
Independently on the incident photon energy, the maximum harmonic energy reaches about 15~eV that agrees with other works \cite{High-Harmonics-semiconductors,Impact}. However, unlike in these references, we consider a wide-band-gap dielectric rather than a semiconductor, and thus we obtain interference patterns between harmonics (straight lines going from the origin of coordinates $\hbar\omega_0 = \hbar\omega$, $\hbar\omega_0 = 2\hbar\omega$ only in two-level system, $\hbar\omega_0 = 3\hbar\omega$, and so on) and polarization generated with photon energy equal to the energy gap $\hbar\omega = E_g$ and corresponding harmonics $\hbar\omega = E_g \pm\hbar\omega_0$, $\hbar\omega = E_g \pm 2\hbar\omega_0$, $\hbar\omega = E_g \pm 3\hbar\omega_0$ and so on. 


\subsection{\label{sec:IV} Electron dynamics}

\begin{figure}[ht!]
\centering
\includegraphics[width=0.99\linewidth]{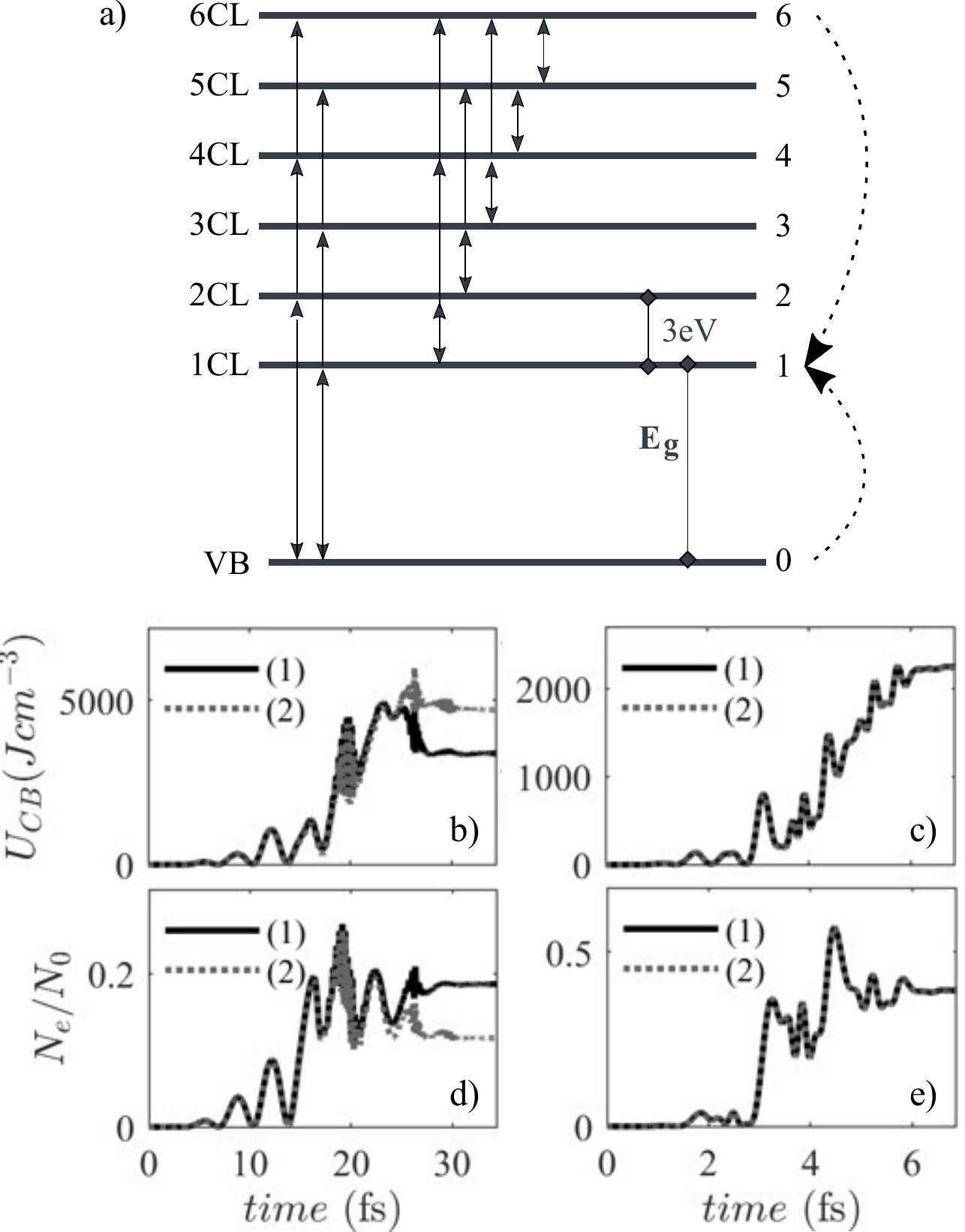}
\caption{\label{fig:3eV_EVO} (a) Illustration of energy levels and transitions between them in the non-centrosymmetric-material OBEs multilevel model. Solid lines indicate field induced transitions. Dotted lines indicate impact ionization pathways. Here the photon energy of the incident light is equal to $\hbar\omega_0=3$~eV, the band gap is $E_g=9$~eV and the corresponding number of considered CB levels is $N = 6$. Time evolution of the energy gained by CB electrons (b,c) and ionization degree (d,e) for $\hbar\omega_0=0.6$~eV (a,b) and 3~eV (c,d), $I_0 = 2\times 10^{14}$~W/cm$^2$, and with (solid curves) and without (dashed curves) impact-ionization and coherence-relaxation processes.}
\end{figure}

In this section we use multi-level systems for modelling of the full laser-induced electron dynamics in the CB of a dielectric material by means of OBEs. The material modeling is done as described in Sec.~\ref{subsec:MODEL:SYMMETRY}, taking $N$ conduction band levels, where $N$ is calculated by Eq.~\eqref{eq:model:N} to allow the impact ionization process. Both, centrosymmetric and non-centrosymmetric material cases are considered. Figure~\ref{fig:3eV_EVO}(a) sketches the example of $\hbar\omega_0=3$~eV giving $N=6$ energy levels in the CB for the latter one.



The time-dependent energy density of the electron subsystem is calculated as follows:
\begin{equation}\label{eq:model:U_CB}
 U_{CB}(t) = \sum_{j>1}^{N}  N_0 (E_j-E_1)\rho_{i,i}(t),
\end{equation}
where $E_1$ is the energy associated to the lowest CB level. It is important to stress that the energy density in the electronic subsystem is calculated for higher CB levels ($j>1$) excluding the lowest CB level ($j=1$), i.e., here we calculate the laser-induced electron heating. 

Figure~\ref{fig:3eV_EVO}(b-e) presents the time evolution of 
the ionization degree $Z(t)=N_e(t)/N_0$, given by Eq.~\eqref{eq:model:Ne}, and the energy density $U_{CB}(t)$ evolution, given by Eq.~\eqref{eq:model:U_CB}, 
for $\hbar\omega_0$ equal to 0.6 and 3.0~eV, $I_0 = 2\times 10^{14}$~W/cm$^2$, and with/without impact-ionization and coherence relaxation processes. For both photon energies, the time evolution of $N_e(t)$ and $U_{CB}(t)$ presents oscillations because electrons go indistinctly to a higher or a lower level. However, this process is not completely reversible. Thus, at the end of the interaction, for $\hbar\omega_0=3$~eV (Fig.~\ref{fig:3eV_EVO} c,e), 
the energy density $U_{CB}$ reaches a value close to the Laser-Induced Damage-Threshold (LIDT) $\sim 2$~kJ/cm$^3$ \cite{Gallais2015} and the electron density in the CB is close to one half of 
initial electron density in VB. 
For this photon energy, effect of impact ionization and coherence relaxation is negligible. 


For $\hbar\omega_0=0.6$~eV (Fig.~\ref{fig:3eV_EVO} b,d), since mid-infrared photons heat better and the absolute laser duration is longer for 5 cycles, including impact ionization and coherence relaxation leads to an increase of the ionization degree but a decrease of the energy density compare to the model without these processes. The final value of $Z$ is approximately one half of the value from $\hbar\omega_0=3.0$~eV. However, the electron heating doubles the ultraviolet case. Ultraviolet photons, even if they ionize more due to their lower multi-photon order, are less efficient to heat electrons in the CB.

\begin{figure}[ht]
\centering
\includegraphics[width=0.85\linewidth]{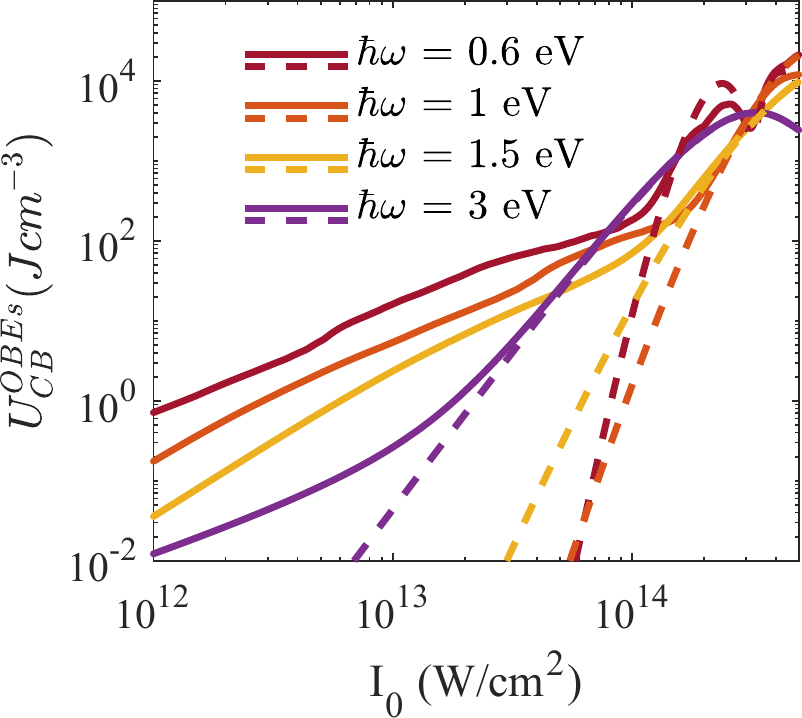}
\caption{\label{fig:Z_U_2L_OBEs}Energy density stored in the non-centosymmetric electron subsystem driven by 5-cycle laser pulse, for several photon energies, with coherence-relaxation and impact ionization processes switched off (dashed curves) and switched on (solid curves).}
\end{figure}

Figure~\ref{fig:Z_U_2L_OBEs} shows 
the energy stored in the CB, $U_{CB}$, driven by 5-cycle 
laser pulse 
in a non-centrosymmetric material with and without impact-ionization and coherence-relaxation processes (in centrosymmetric materials similar results are obtained and thus are not shown). 
For low intensities $I_0<10^{13}$~W/cm$^{2}$ the relaxation and impact ionization terms lead to a change of the $Z$ (not shown here) and $U_{CB}$ dependence on the incident pulse intensity $I_0$. The lower the photon energy $\hbar\omega_0$, the stronger the departure from the expected multi-photon rate at low intensities and the stronger the influence of the coherence-relaxation and impact ionization terms. 

For high intensities $I_0 > 5\times 10^{14}$~W/cm$^{2}$, instead, the energy density gained by CB electrons is similar in both cases because the dielectric material is fully ionized. Between these two intensity limits, impact-ionization and coherence-relaxation processes play a significant role in the Mid-Infrared region ($\hbar\omega_0<1$~eV) because the order of the multi-photon ionization process is higher. 

\section{\label{sec:Conclusion} Conclusion}
In order to describe the time-dependent electron dynamics in dielectric materials induced by ultra-short few-cycle laser pulses, a modeling based on optical Bloch equations has been developed. Through the introduction of an appropriate set of energy levels mimicking the band structure of dielectric materials, the present approach includes a description of photo-ionization, electron heating in the conduction band, impact ionization, collisions in the conduction band, and electron recombination. The reliability of this approach has been assessed by studying various physical quantities. First, the evolution of the density of the conduction electrons with respect to the laser intensity in case of pure photo-ionization has been studied. By changing the wavelength, we have shown that this model is able to account for the multiphoton absorption in the perturbative case, i.e. not too high intensities. For intensities in excess of $10^{14} W/cm^2$, the tunneling regime is entered leading to a saturation of the ionization rate with respect to the intensity. Second, the evaluation of the polarization has been provided. Its Fourier transformation confirms that harmonic generation is well described up to high orders. In particular, it is demonstrated that properties of symmetry of the material lattice can be included: only odd harmonics are generated by imposing a centro-symmetric structure. Finally, the full electron dynamics has been studied. The temporal evolution of the density of conduction electrons clearly exhibits the ability of this modeling to describe the time-dependent electron dynamics driven the oscillating electric laser field. The electron energy density is shown to follow such a behavior. The influence of the impact ionization is also observed. As expected, the lower the photon energy, the larger its contribution.
\par
Overall, the present modeling describes all expected behaviors regarding the laser-induced electron dynamics in case where all main collisional processes are included. This approach thus provides a theoretical baseline well adapted to describe accurately the electron dynamics driven by few-cycle laser pulse. Such an approach is well adapted to be introduced in a Maxwell solver to describe the coupled electron-pulse propagation dynamics. Such a coupling will allow one to provide accurate predictions of the energy deposition in dielectric materials. It may thus provide a step forward for numerically designing experiments and applications.


\section{Acknowledgements}
Numerical simulations were performed using computing resources at Grand
{\'E}quipement National pour le Calcul Intensif (GENCI, Grants
No.~A0030506129 and No.~A0040507594) and Chalmers Centre for
Computational Science and Engineering (C3SE) provided by the Swedish
National Infrastructure for Computing (SNIC, Grant SNIC 2018/4-38).

\appendix



\section{\label{sec:appendix:numerics} Numerical resolution of OBEs}

In the context of Finite-Difference Time-Domain (FDTD) Maxwell-Bloch simulations, the equations of our OBEs model are coupled to a Maxwell solver, e.g., the Yee scheme \cite{Yee1966}. In this section, we present the numerical scheme used in this paper to solve OBEs equations and which can be easily integrated into such FDTD algorithms:
\begin{equation}\label{eq:appendix:numerics:OBEs}
\partial_t{\hat{\rho}} = \hat{\cal L}\left(\hat{\rho} \right) + \hat{\cal G}_{\rm rec}\left(\hat{\rho} \right) + \hat{\cal G}_{\rm coh}\left(\hat{\rho} \right) + \hat{\cal G}_{\rm imp}\left(\hat{\rho} \right).
\end{equation}

Let us consider that the electric field $E(t)$ is known, in general from the Maxwell solver and particularly in this paper from Eq.~\eqref{eq:model:laser}, at the discrete instants $\left\lbrace t_n = n \Delta t  \right\rbrace$, where $\Delta t$ is the time step and $n\geq 0$. This time instants will constitute the {\it primal} temporal grid. At these primal instants the Hamiltonian is fully known according to Eqs.~\eqref{eq:model:H}~and~\eqref{eq:model:OBEs:V}:
\begin{equation}\label{eq:appendix:numerics:H tn}
\hat{H}(t_{n})  =  \hat{H}_0 - e E(t_n) \hat{\mu}.
\end{equation}

FDTD codes usually update alternatively the electric field $E(t)$ and material response variables (i.e., the electron density $N_{\rm e}(t)$ and the current density $J(t)$). In consequence, we need to construct a numerical scheme allowing us to compute $\hat{\rho}$ at the discrete instants $\left\lbrace t_{n-1/2}=t_n-\Delta t/2 \right\rbrace$, which will constitute the {\it dual} temporal grid. All the material response variables can be computed from the density matrix. We shall use the notation $\hat{\rho}_{n-1/2}$ for the numerical computation of the density matrix at dual instant $t_{n-1/2}$. For centro-symmetric material (for details see Section~\ref{subsec:MODEL:SYMMETRY}), we impose as initial condition that all the two VB levels are equally populated and that all the $N$ CB levels are completely depleted:
\begin{equation}\label{eq:appendix:numerics:initial condition}
\hat{\rho}_{-1/2} = \mqty[\dmat[0]{\frac{1}{2},\frac{1}{2},0,\ddots,0}].
\end{equation}

During each update step, in order to calculate $\hat{\rho}_{n+1/2}$ from $\hat{\rho}_{n-1/2}$ and $\hat{H}(t_n)$, each of the super-operators in Eq.~\eqref{eq:appendix:numerics:OBEs} is treated separately following the Strang splitting approach \cite{Strang1968}. We seek a second-order accurate scheme at every iteration and, since the the numerical cost of applying the relaxation super-operators is considerably smaller than applying the Liouville-von-Neumman super-operator, we update the density matrix in seven sub-steps as follows:
\begin{equation}\label{eq:appendix:numerics:strang}
\left\lbrace \begin{array}{l}
     \hat{\rho}_{n+1/2}^{(1)} = \hat{G}^{\Delta t/2}_{\rm rec} \left( \hat{\rho}_{n-1/2} \right), \\
     \\
     \hat{\rho}_{n+1/2}^{(2)} = \hat{G}^{\Delta t/2}_{\rm imp} \left( \hat{\rho}_{n+1/2}^{(1)} \right) ,\\
     \\
     \hat{\rho}_{n+1/2}^{(3)} = \hat{G}^{\Delta t/2}_{\rm coh} \left( \hat{\rho}_{n+1/2}^{(2)} \right), \\
     \\
     \hat{\rho}_{n+1/2}^{(4)} = \hat{L}^{\Delta t} \left( \hat{\rho}_{n+1/2}^{(3)}, \, \hat{H}(t_n) \right), \\
     \\
     \hat{\rho}_{n+1/2}^{(5)} = \hat{G}^{\Delta t/2}_{\rm coh} \left( \hat{\rho}_{n+1/2}^{(4)} \right) ,\\
     \\
     \hat{\rho}_{n+1/2}^{(6)} = \hat{G}^{\Delta t/2}_{\rm imp} \left( \hat{\rho}_{n+1/2}^{(5)} \right), \\
     \\
     \hat{\rho}_{n+1/2} = \hat{G}^{\Delta t/2}_{\rm rec} \left( \hat{\rho}_{n+1/2}^{(6)} \right),
\end{array}       \right .
\end{equation}
where $ \hat{G}^{\Delta t}_{\rm rec} $ is the discrete recombination super-operator acting over $\Delta t$, $ \hat{G}^{\Delta t}_{\rm imp} $ is the discrete impact-ionization super-operator acting over $\Delta t$, $ \hat{G}^{\Delta t}_{\rm coh} $ is the discrete coherence-loss super-operator acting over $\Delta t$, and $ \hat{L}^{\Delta t} $ is the discrete Liouville-von-Neumann super-operator acting over $\Delta t$. In order to preserve the properties of the density matrix over the simulation time span, the discretization of the ensemble of super-operators during each update step must constitute a Completely Positive Trace Preserving (CPTP) map \cite{Lindblad1966,Gorini1976}. In order to have a CPTP splitting in Eq.~\eqref{eq:appendix:numerics:strang} and thus obtain numerical solutions compatible with physics, we must assure that all the discrete super-operators (namely, $ \hat{G}^{\Delta t}_{\rm rec} $, $ \hat{G}^{\Delta t}_{\rm imp} $, $ \hat{G}^{\Delta t}_{\rm coh} $ and $ \hat{L}^{\Delta t} $) are CPTP \cite{Riesch2018}.

In case of two VB levels, the recombination super-operator introduces recombination process from each CB level to each VB level (k = 0, k = -1) with the same characteristic time scale $\tau_{rec}$ similarly to the case of one VB (Eq.~\eqref{eq:model:G rec i}) and is discretized as follows:
\begin{equation}\label{eq:appendix:numerics:G rec sum}
\hat{G}^{\Delta t}_{\rm rec}\left( \hat{\rho} \right) = \sum_{k = -1}^{0} \hat{G}^{\Delta t}_{\rm rec \ k}\left( \hat{\rho} \right),
\end{equation}
where
\begin{equation}\label{eq:appendix:numerics:G rec}
\hat{G}^{\Delta t}_{\rm rec \ k}\left( \hat{\rho} \right) \equiv \left\lbrace    \begin{array}{ll}
     \displaystyle  \rho_{j,j} =  \rho_{j,j} - \Delta \rho_{j,j}, & \text{for $j>0$,} \\
     \\
     \displaystyle  \rho_{k,k} = \rho_{k,k} + \sum_{j=1}^N  \Delta \rho_{j,j}
      
\end{array}  \right .
\end{equation}
where $\Delta \rho_{j,j} = ( 1 - {\rm e}^{ - \Delta t/\tau_{rec} }) \rho_{j,j}$ is the decay in the electron population in the $j$-th CB level over the time $\Delta t$ due to recombination transition to the $k$-th VB level. It is straightforward to verify that this discretization of the recombination super-operator conserves the trace and the positivity of the diagonal elements of $\hat{\rho}$.

Following Eq.~\eqref{eq:model:G imp (k)}, the impact-ionization super-operator acting over a time $\Delta t$ on a system with two VB levels is discretized as follows:
\begin{equation}\label{eq:appendix:numerics:G imp}
\hat{G}_{\rm imp}^{\Delta t}\left(\hat{\rho} \right) \equiv \left \lbrace  \begin{array}{ll}
     \displaystyle  \rho_{N,N} =  \rho_{N,N} - \sum_{k=-1}^0  \Delta \rho_{k,k},   \\
      \\
     \displaystyle \rho_{1,1} = \rho_{1,1} + \sum_{k=-1}^0  2 \, \Delta \rho_{k,k},   \\
      \\
    \displaystyle  \rho_{k,k} =  \rho_{k,k} - \Delta \rho_{k,k} , & \text{for $k\leq 0$,}
\end{array}   \right .
\end{equation}
where, here, $\Delta \rho_{k,k} $ represents the decay in the electron population of the $k$-th VB level and is given by:
\begin{equation}\label{eq:appendix:numerics:G imp:Delta rho kk}
\Delta \rho_{k,k} = \min \left\lbrace \rho_{k,k}, \, \left(1-{\rm e}^{-\Delta t/\tau_{imp} }\right)\rho_{N,N} \right\rbrace.
\end{equation}

The introduction of the min-function in Eq.~\eqref{eq:appendix:numerics:G imp:Delta rho kk} when computing $\Delta \rho_{k,k}$ ensures the preservation of the non-negativeness of all VB levels: $\rho_{k,k} \geq 0$ for $-1\leq k \leq 0$.

The discretization of the coherence-loss super-operator, given by Eq.~\eqref{eq:model:G coh}, acting over a time $\Delta t$, is the following:
\begin{equation}\label{eq:appendix:numerics:G coh}
\hat{G}_{\rm coh}^{\Delta t}\left(\hat{\rho} \right) \equiv \left \lbrace  \begin{array}{ll}
    \rho_{j,k} = {\rm e}^{- \Delta t/\tau_{coh} } \, \rho_{j,k}, & \text{for $j\neq k$,} \\
\end{array}   \right .
\end{equation}
which preserves all the properties of density matrix $\hat{\rho}$.

There are several possibilities of obtaining a CPTP discrete Liouville-von-Neumann super-operator, such as the Crank-Nicolson approach \cite{Ziolkowski1995,Slavcheva2002}, Runge-Kutta methods \cite{Sukharev2011, Deinega2014, Cartar2017} and the matrix exponential approach \cite{Bourgeade2012,Riesch2018}. The latter technique is chosen in this paper because it adapts well to the alternatively-updating nature of FDTD codes. Assuming that during the time interval $[t_{n-1/2}, t_{n+1/2}]$ the Hamiltonian in the Liouville-von-Neumann super-operator is time-independent and equal to $\hat{H}(t_n)$ given by Eq.~\eqref{eq:appendix:numerics:H tn}, then the exact solution to $\partial_t \hat{\rho} = \hat{\cal L}(\hat{\rho})$ reads:
\begin{equation}\label{eq:appendix:numerics:sol L}
\hat{\rho}(t) = \hat{\cal H}(t) \, \hat{\rho}_{n-1/2} \, \hat{\cal H}(t)^\dagger,
\end{equation}
where the $\hat{\cal H}(t)$ is the following matrix exponential:
\begin{equation}\label{eq:appendix:numerics:sol L:ME}
\hat{\cal H}(t) = \exp\left[ -\frac{\rm i}{\hbar}\,\hat{H}(t_n) \, (t-t_{n-1/2}) \right].
\end{equation}

Since $H(t_n)$ is real and symmetric in our paper, $\hat{\cal H}(t)^\dagger$ reduces to $\hat{\cal H}(t)^*$ in Eq.~\eqref{eq:appendix:numerics:sol L}. Therefore, discretization of the Liouville-von-Neumann super-operator in Eq.~\eqref{eq:appendix:numerics:strang} reads as follows:
\begin{equation}\label{eq:appendix:numerics:L}
\hat{L}^{\Delta t}\left(\hat{\rho}, \hat{H} \right) \equiv \left\lbrace \begin{array}{l}
    \displaystyle \hat{\rho} = \hat{\cal H}  \, \hat{\rho} \, \hat{\cal H} ^*,  \\
\end{array}   \right.
\end{equation}
where
\begin{equation}\label{eq:appendix:numerics:L:H}
\hat{\cal H} =\hat{\cal H}(\Delta t, \hat{H})  =\exp\left[-\frac{\rm i}{\hbar}\hat{H}\Delta t\right],
\end{equation}
which is CPTP because it is an exact solution of the Liouville-von-Neumann equation. In practice, since the numerical calculation of matrix exponentials requires high computational ressources \cite{Moler2003}, we can use a second-order accurate-in-time approximation of Eq.~\eqref{eq:appendix:numerics:L:H} provided that the norm $\norm{-{\rm i}\hat{H}\Delta t/ \hbar}\ll 1$ for a time step $\Delta t$ being sufficiently small. The approximation that we use in this paper, which is CPTP, reads as follows \cite{BidegarayFesquet2006, Bourgeade2012}:
\begin{equation}\label{eq:appendix:numerics:L:H:approx}
\exp\left[-\frac{\rm i}{\hbar}\hat{H}\Delta t\right] \approx \hat{\cal I} \left(  \hat{\cal I} ^* \right)^{-1} + \order{\Delta t^3} ,
\end{equation}
where:
\begin{equation}\label{eq:appendix:numerics:L:H:approx:I}
\hat{\cal I} = \hat{I} - \frac{{\rm i}\Delta t}{2\hbar} \hat{H},
\end{equation}
and $\hat{I}$ accounts for the identity matrix. Thanks to the approximation~\eqref{eq:appendix:numerics:L:H:approx} only one matrix inversion and two matrix conjugations are computed at each time iteration. We employed the library LAPACK \cite{laug} to do so.

Finally, we compute the material response variables from the density matrix at the dual instants. The electron density can be easily calculated thanks to  Eq.~\eqref{eq:model:Ne}:
\begin{equation}\label{eq:appendix:numerics:Ne}
N_{\rm e}(t_{n+1/2}) = N_0 \, \sum_{j>0}^N \rho_{j,j}(t_{n+1/2}) 
\end{equation}
And the polarization response is calculated accordingly to Eq.~\eqref{eq:model:OBEs:P}:
\begin{equation}\label{eq:appendix:numerics:P}
P(t_{n+1/2}) = N_0  \Tr\left\lbrace e\hat{\mu}\,\left(\hat{\rho}(t_{n+1/2}) -\hat{\rho}(0) \right) \right\rbrace
\end{equation}

The current density is then computed as a time derivative of the polarization response :
\begin{equation}\label{eq:appendix:numerics:J}
\begin{multlined}
J(t_{n+1/2}) =  N_0   \Tr \left\lbrace    -\frac{\rm i}{\hbar}e\hat{\mu}\comm{\frac{ \hat{H}(t_n)+\hat{H}(t_{n+1})}{2}}{\hat{\rho}_{n+1/2}}\,\right. \\
+ ~ e \hat{\mu}\hat{\cal G}_{\rm rec}\left(\hat{\rho} _{n+1/2}\right) 
\left.~ +~ e \hat{\mu}\hat{\cal G}_{\rm coh}\left(\hat{\rho} _{n+1/2}\right)\right. \\ 
\left.~ +~e \hat{\mu}\hat{\cal G}_{\rm imp}\left(\hat{\rho}_{n+1/2}\right) \right\rbrace 
\end{multlined}
\end{equation}

\bibliography{biblio.bib}

\end{document}